\pdfoutput=1
\documentclass[useAMS,usenatbib]{mnras}
\usepackage{graphicx,aas_macros}
\usepackage{url}
\usepackage{color}
\usepackage{bm} 
\usepackage{amssymb}
\usepackage{amsbsy}
\usepackage{amsmath}
\usepackage{hyperref}
\usepackage{times}
\usepackage{rotating}
\usepackage[usenames,dvipsnames]{xcolor}
\usepackage[colorinlistoftodos]{todonotes}

\title[Bayesian constraints on the global 21-cm signal]{Bayesian constraints on the global 21-cm signal from the Cosmic Dawn}

\hypersetup{
    breaklinks=true,
    unicode=true,                     
    linktocpage=true,
    pdftoolbar=true,                
   pdfmenubar=true,              
    pdffitwindow=true,           
    pdfstartview={FitH},        
    pdfhighlight={/I},
   hyperindex=true,
    pdftitle={My title},            
   pdfauthor={Jonathan Zwart},         
   pdfsubject={Subject},       
   pdfcreator={Creator},       
    pdfproducer={Producer}, 
    pdfkeywords={keywords}, 
    pdfnewwindow=true,      
   colorlinks=true,       
    linkcolor=red,          
    citecolor=blue,        
    filecolor=magenta,      
    urlcolor=cyan           
}

\author[Bernardi~et~al.]{G.~Bernardi\thanks{gbernardi@ska.ac.za}$^{1,2,3}$,
  J.~T.~L.~Zwart$^{4,5}$, D.~Price$^{3}$, L.J.~Greenhill$^{3}$, A.~Mesinger$^6$, J.~Dowell$^{7}$, \newauthor T.~Eftekhari$^{3}$, S.W.~Ellingson$^{8}$, J.~Kocz$^9$ and F.~Schinzel$^7$\\
$^1$SKA South Africa, 3rd Floor, The Park, Park Road, Pinelands 7405, South Africa\\
$^2$ Department of Physics and Electronics, Rhodes University, PO Box 94, Grahamstown 6140, South Africa\\
$^3$Harvard-Smithsonian Center for Astrophysics, Garden Street 60, Cambridge, MA, 02138, USA\\
$^4$Department of Physics \& Astronomy, University of the Western Cape, Private Bag X17, Bellville, Cape Town 7535, South Africa\\
$^5$Astrophysics, Cosmology \& Gravity Centre, Department of Astronomy, University of Cape Town, Private Bag X3, Rondebosch 7701, South Africa\\
$^6$Scuola Normale Superiore, Piazza dei Cavalieri 7, 56126 Pisa, PI, Italy\\
$^7$Department of Physics and Astronomy, University of New Mexico, Albuquerque, NM 87131, USA\\
$^8$Bradley Dept. of Electrical \& Computer Engineering, Virginia Tech, Blacksburg VA 24061, USA\\
$^9$Jet Propulsion Laboratory, 4800 Oak Grove Drive, Pasadena, CA, 91104, USA
}

\begin{document}
\date{Accepted ---. Received ---; in original form \today.}

\pagerange{\pageref{firstpage}--\pageref{lastpage}} \pubyear{2016}

\label{firstpage}

\maketitle

\begin{abstract}
The birth of the first luminous sources and the ensuing epoch of reionization are best studied via the redshifted 21-cm emission line, the signature of the first two imprinting the last. In this work we present a fully-Bayesian method, \textsc{hibayes}, for extracting the faint, global (sky-averaged) 21-cm signal from the much brighter foreground emission. We show that a simplified (but plausible), Gaussian model of the 21-cm emission from the Cosmic Dawn epoch ($15 \lesssim z \lesssim 30$), parameterized by an amplitude $A_{\rm HI}$, a frequency peak $\nu_{\rm HI}$ and a width $\sigma_{\rm HI}$, can be extracted even in the presence of a structured foreground frequency spectrum (parameterized as a $7^{\rm th}$-order polynomial), provided sufficient signal-to-noise (400~hours of observation with a single dipole). We apply our method to an early, 19-minute long observation from the Large aperture Experiment to detect the Dark Ages, constraining the 21-cm signal amplitude and width to be $-890 < A_{\rm HI} < 0$~mK and $\sigma_{\rm HI} > 6.5$~MHz (corresponding to $\Delta z > 1.9$ at redshift $z \simeq 20$) respectively at the 95-per-cent confidence level in the range $13.2 < z < 27.4$ ($100 > \nu > 50$~MHz).
\end{abstract}


\section{Introduction}
\label{sec:intro}
Observations of the highly-redshifted 21-cm emission are considered
the most powerful probe of the birth of the first luminous sources and
the consequent epoch of reionization \citep[for recent reviews
see][]{mcquinn15,furlanetto15}. The 21-cm emission precisely traces and times the evolution of the average Hydrogen neutral fraction and the growth of the HII regions around ionizing sources throughout reionization \citep[e.g.,][]{ciardi03,mellema06,mcquinn07,lidz08}. 

Prior to reionization, during the so-called Cosmic Dawn, the 21-cm signal marks the Ly$\alpha$ coupling and the X-ray heating eras respectively. The Ly$\alpha$ coupling occurs with the birth of the first luminous sources that are expected to be highly effective in coupling the spin temperature to the InterGalactic Medium (IGM) temperature, generating 21-cm emission via the Wouthuysen--Field effect \citep[]{wouthuysen52,field59}. Most models anticipate this transition to occur around $z \simeq 25-30$, when the IGM is colder than the Cosmic Microwave Background (CMB), causing a 21-cm signal in absorption against the CMB. The 21-cm emission is here sensitive to the nature of the first luminous sources as well as the details of the formation of the first galaxies in the first minihalos \citep[e.g.,][]{ciardi03b,furlanetto06,fialkov13}. In particular, the redshift and amplitude of the peak in the 21-cm emission strongly depends on whether the dominant contribution to the Ly$\alpha$ coupling comes from atomically-cooled galaxies or minihalos, and how much the Lyman--Werner background suppresses star formation \citep[e.g.,][]{haiman00,ricotti01,fialkov13}. 

As star formation progresses, X-rays are generated in the first galaxies by either early black holes or the diffuse, hot interstellar medium. Although other sources of energy injection due to dark matter annihilation \citep[]{valdes07,valdes13} and shocks from fluid motions may be present in the early Universe \citep{mcquinn12}, X-ray emission is commonly believed to be the most significant source of IGM heating that would, eventually, drive its temperature above the CMB temperature \citep[]{pritchard07,mesinger13,pacucci14,tanaka16}. The relative timing of this process is, however, very uncertain due to the essentially unknown properties of the first galaxies \citep{mesinger16}. In particular, most models assume that the IGM is heated well above the CMB temperature by the onset of reionization. However, if the first galaxies show a hard X-ray spectrum or their X-ray efficiency (commonly parameterized as the number of X-ray photons produced per stellar baryon) is low, then heating becomes inefficient and reionization begins when the IGM is still colder than the CMB \citep[the `cold reionization' scenario,][]{fialkov14,mesinger14}. Such a scenario would also impact the subsequent morphology of reionization \citep{iliev12,ewall-wice16b}.

This theoretical landscape is still completely unconstrained by observations, but first upper limits to the 21-cm fluctuations in the $12 < z < 18$ range are starting to appear at approximately three orders of magnitude higher than the expected signal \citep{ewall-wice16a}. There is also initial evidence of heating prior to reionization provided by recent 21-cm power spectrum upper limits at $z = 8.4$ that constrain the IGM to be warmer than 8~K \citep[]{ali15,pober15,greig16}.

While current experiments targeting 21-cm fluctuations are well placed to constrain the reionization process statistically, only the upcoming interferometric arrays like the Hydrogen Epoch of Reionization Array \citep[]{pober14,deboer16} and the Square Kilometre Array \citep{koopmans15} will have sufficient sensitivity and frequency coverage to probe the Ly$\alpha$ and X-ray heating epochs \citep{mesinger15,ewall-wice16b}. Therefore increased attention has recently been devoted to observations targeting the global (sky-averaged) 21-cm emission \citep[e.g.,][]{pritchard10,morandi12,liu16}, including novel ways to use interferometric arrays to probe the global 21-cm signal \citep[][]{mckinley13,presley15,sing15,vedantham15}. Albeit challenged by the same requirements of accurate subtraction of bright foreground emission and control over systematic effects that affect its sibling 21-cm fluctuation observations, the global 21-cm signal may represent an alternative, relatively inexpensive way to achieve the milliKelvin sensitivity needed to access the pre-reionization epoch. Instruments like the Experiment to Detect the Global
Epoch-of-Reionization Signature \citep[EDGES,][]{bowman08,bowman10}, the Large aperture Experiment to detect the
Dark Ages \citep[LEDA,][]{greenhill12,bernardi15,kocz15,price2016}, SCI--HI \citep{voytek14} and the Dark Age Radio Explorer \citep[DARE,][]{mirocha15,harker16} are (or will be) targeting such an epoch.

In this paper we present a Bayesian foreground separation method and show that it can extract the 21-cm signal from the Cosmic Dawn even in the presence of non--spectrally-smooth foreground emission parameterized through high-order polynomials in frequency. We apply the algorithm to early LEDA data to derive upper limits on the global 21-cm signal in the $13.2 < z < 27.4$ range.

The paper is organized as follows. In Section~\ref{sec:bayes} we describe the Bayesian method and its application to simulated data. In Section~\ref{sec:observations}, we apply it to LEDA data; our results are discussed in Section~\ref{sec:conclusions}.

\section{Bayesian framework and simulations}
\label{sec:bayes}
\begin{figure}
\centering
\includegraphics[width=1.\columnwidth]{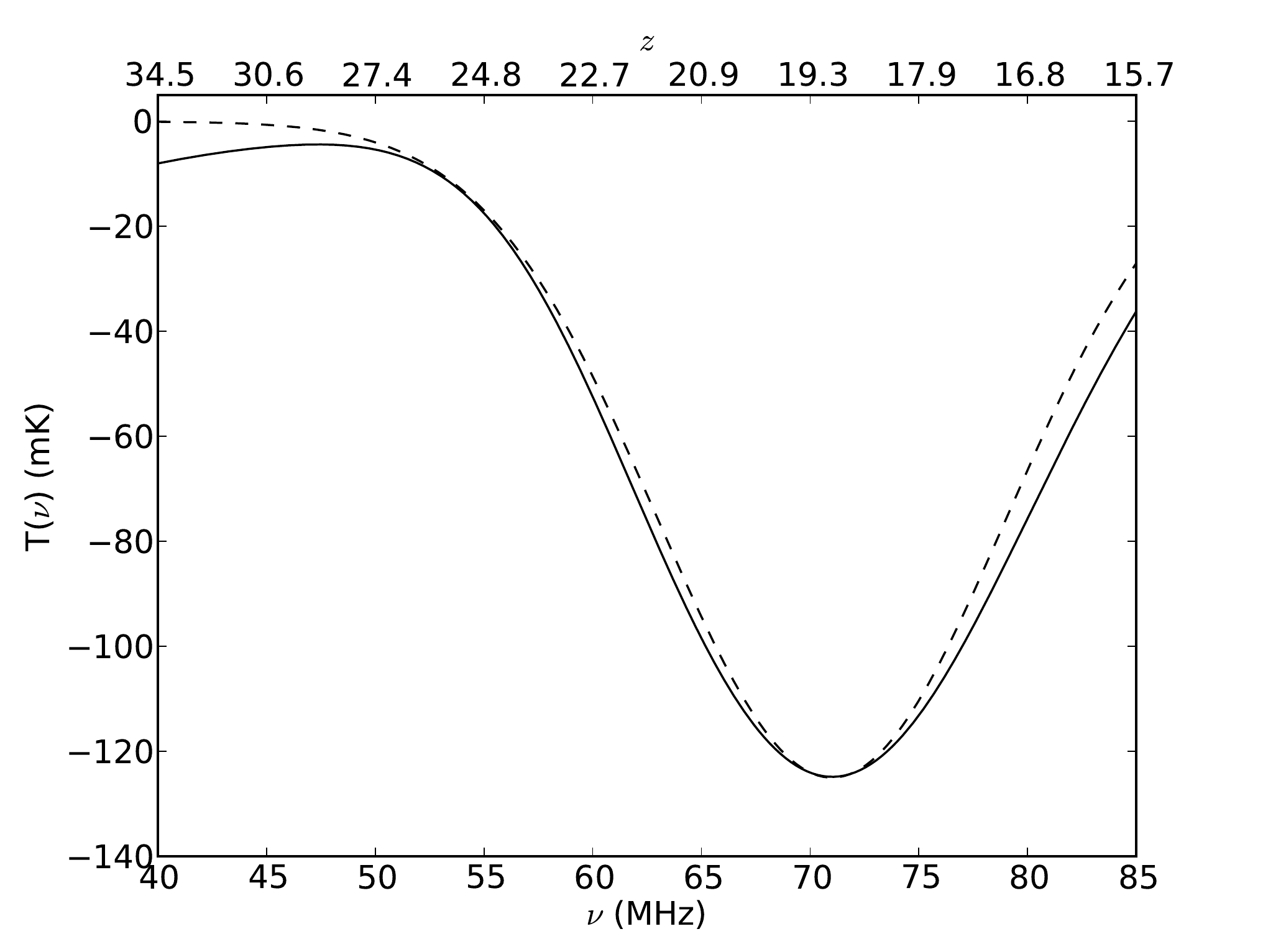}
\caption{Comparison between a 21-cm empirical Gaussian model (dashed line) used in our current analysis and a physical model derived from ARES (solid line). The physical model is taken from \citet{mirocha15} and is defined by four parameters: the minimum virial temperature for star-forming halos $T_{\rm min}$, the efficiencies of Ly$\alpha$ and X-ray photon production, $\xi_{\rm LW}$ and $\xi_{\rm X}$ respectively, and the IGM ionization efficiency $\xi_{\rm ion}$. We set $T_{\rm min} = 10^4$~K, $\xi_{\rm LW} = 969$, $\xi_{\rm X} = 0.02$ and $\xi_{\rm ion} = 40$ respectively \citep[see][for details]{mirocha15} and this can be considered as a reference model. The Gaussian model parameters are $A_{\rm HI} = -125$~mK, $\nu_{\rm HI} = 71$~MHz and $\sigma_{\rm HI} = 8$~MHz, similar to the model chosen for our simulations. The agreement between the two profiles is at the 10--20-per-cent level across most of the LEDA band.}\label{fig:model_comparison}
\end{figure}
\begin{figure*}
\centering
\includegraphics[width=1.0\textwidth]{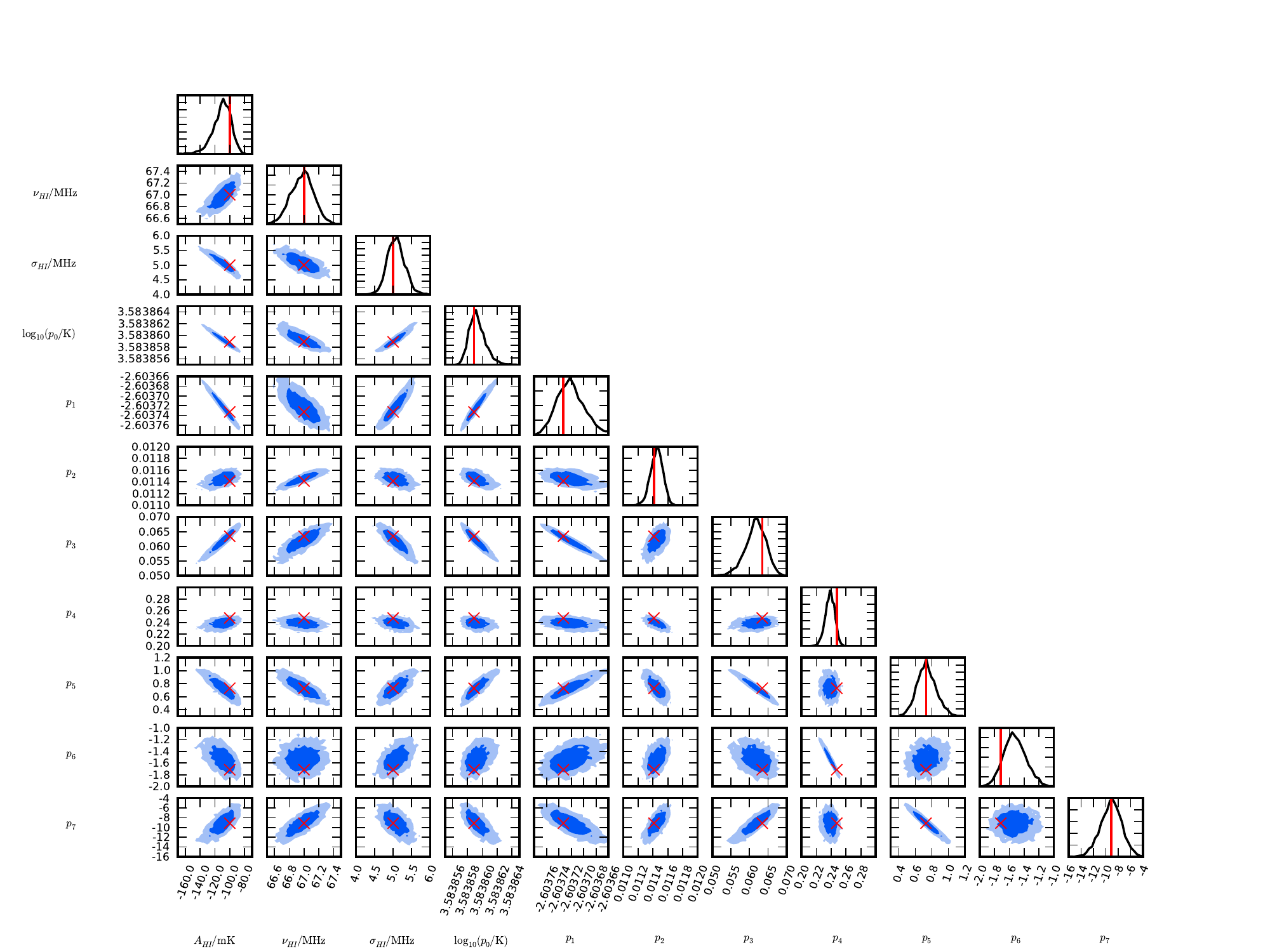}
\caption{Posterior probability distribution, marginalized into one and
  two dimensions, for the $N=7$ foreground and the 21-cm models fitted
  to the simulated data. The dark and light shaded regions indicate
  the 68- and 95-per-cent confidence regions. The simulated parameter values are indicated in red. The marginalized probability distributions are plotted in the $[0,1]$ range.}\label{fig:triangle_sim_plot}
\end{figure*}
\begin{figure*}
\centering
\includegraphics[width=1.0\textwidth]{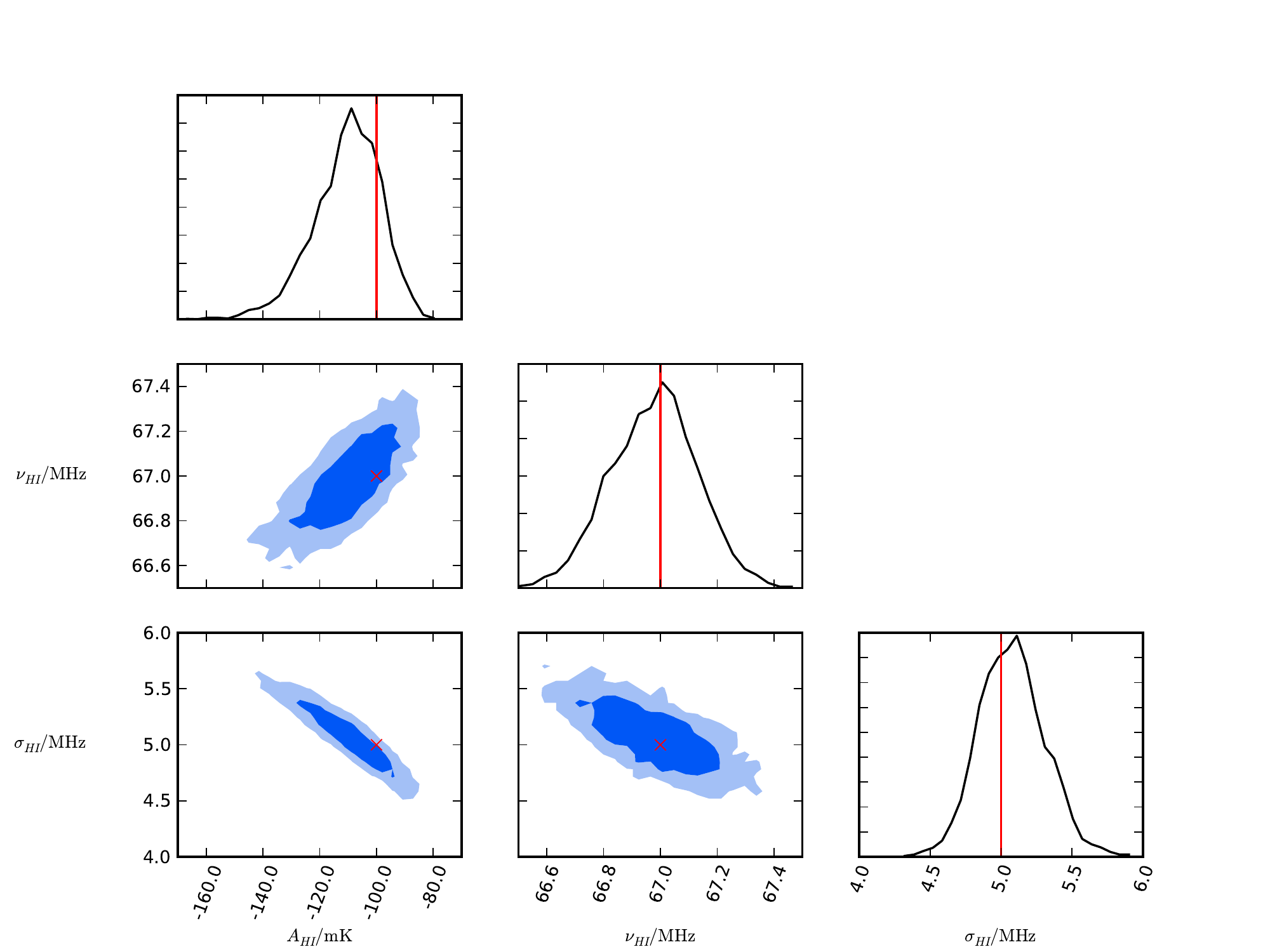}
\caption{Zoom-in on the 21-cm parameters from
  Figure~\ref{fig:triangle_sim_plot}. The dark and light shaded regions indicate
  the 68- and 95-per-cent confidence regions. The simulated parameter values
  are indicated in red. The marginalized probability distributions are plotted in the $[0,1]$ range.}
\label{fig:triangle_sim_zoomin_plot}
\end{figure*}

Bayesian Monte Carlo sampling  (for example using Markov Chains; MCMC) has become a
standard method for exploring a likelihood surface and reconstructing the
posterior distribution in order to extract cosmological parameters
from CMB observations and, recently, also in the 21-cm field
\citep[]{harker12,greig15}.
Bayes' theorem indeed relates the posterior probability distribution
$\mathcal{P}({\bf \Theta} | {\bf D}, \mathcal{H})$ of a set of
parameters ${\bf \Theta}$ given the data ${\bf D}$ and a model
${\mathcal H}$, that includes the hypothesis and any related
assumptions, to the likelihood
$\mathcal{L}({\bf D} | {\bf \Theta}, \mathcal{H})$ as:
\begin{equation}
\label{eqn:bayes}
\mathcal{P}\left(\mathbf{\Theta}|\mathrm{\mathbf{D}},\mathcal{H}\right)
= \frac{
\mathcal{L}\left(\mathrm{\mathbf{D}}|\mathbf{\Theta},\mathcal{H}\right)
\mathit{\Pi}\left(\mathbf{\Theta}| \mathcal{H}\right)}
{\mathcal{Z}\left(\mathrm{\mathbf{D}}| \mathcal{H}\right)},
\end{equation}
\noindent
where the priors
${\mathit \Pi}\left(\mathbf{\Theta}|\mathcal{H}\right)$ encode
existing knowledge of parameter values and the evidence
${\mathcal Z}\left(\mathrm{\mathbf{D}}| \mathcal{H}\right)$ is the
integral of the likelihood
${\mathcal L}({\bf D} | {\bf \Theta}, {\mathcal H})$ over the prior
space, allowing not only normalization of the posterior but also model
selection via its inherent ability to quantify Occam's razor \citep[e.g.][]{mackay03,liddle06,trotta08,parkinson13}.

We implemented an algorithm for extracting the global 21-cm signal
following \cite{harker12}, who assume Gaussian measurement noise and
hence write the likelihood $\mathcal{L}_j$ of measuring the observed
sky temperature $T_{\rm ant}(\nu_j)$ at a single frequency $\nu_j$ as:
\begin{equation}
{\mathcal L}_j\left(T_{\rm ant} (\nu_j) | {\bf \Theta}\right) = \frac{1} {\sqrt{2 \pi \sigma^2(\nu_j)}} \mathrm{e}^{-\frac{[T_{\rm ant}(\nu_j) - T_m(\nu_j,\mathbf{\Theta})]^2}{2 \sigma^2(\nu_j)}},
\end{equation}
\noindent
where $T_m(\nu_j,\mathbf{\Theta})$ is the model spectrum and
$\sigma(\nu_j)$ is the standard deviation of the frequency-dependent
instrumental noise,
\begin{eqnarray}
\sigma(\nu_j) = \frac{T_{\rm ant} (\nu_j)}{\sqrt{\Delta \nu \Delta t}}, \nonumber
\end{eqnarray}
where $\Delta \nu$ is the channel width and $\Delta t$ is the total integration time.
Assuming that $T_{\rm ant}(\nu)$ is measured at $M$ discrete frequency
channels and that the noise is uncorrelated between frequency
channels, the (log-)likelihood for the full frequency spectrum
becomes:
\begin{equation}
{\ln\mathcal L}\left({\bf T_{\rm ant}} | {\bf \Theta}\right) = \sum_{j=1}^M {\ln\mathcal L}_j\left(T_{\rm ant}(\nu_j) | {\bf \Theta}\right).
\end{equation}
\noindent
The sky model at each frequency channel $\nu_j$ is the sum of the foreground $T_f$ and the 21-cm signal $T_{\rm HI}$:
\begin{eqnarray}
T_m (\nu_j) = T_f (\nu_j) + T_{\rm HI} (\nu_j).
\end{eqnarray}
Single-dipole observations measure the integrated Galactic foreground spectrum averaged over the whole sky, losing information about its spatial structure and how to separate foregrounds from the 21-cm global signal is still a very active debate in the community. \cite{liu13} and \cite{switzer14} suggest taking advantage of the spatial structure of the Galactic foreground in order to improve its separation from the spatially-constant 21-cm global signal. The most commonly adopted approach is to simply leverage the different spectral behaviour of foregrounds and the 21-cm signal, parameterizing the foreground spectrum through a principal component analysis \citep[e.g.][]{vedantham14} or a log-polynomial \citep[e.g.][]{pritchard10,bowman10,harker12,voytek14,bernardi15,presley15}. In this paper we therefore model the foreground emission as a $N^{\rm th}$ order log-polynomial:
\begin{equation}
\log_{10} T_f (\nu_j) = \sum_{n=0}^N p_n \left[ \log_{10} \left( \frac{\nu_j}{\nu_0} \right) \right]^n,
\end{equation}
where we have adopted the convention $\nu_0 = 60$\,MHz.

The choice of the polynomial order is critical in order to correctly model the foreground spectrum. Although earlier works showed that the foreground spectrum can be well described by very few components in frequency \citep[e.g.][]{de-oliveira-costa08,pritchard10}, more recent simulations suggest that most of the frequency structure present in the observed sky arises from the coupling between the sky and the antenna beam pattern \citep[]{bernardi15,mozdzen15}. 

Our implementation is focused on the pre-reionization, Cosmic Dawn signal at
$15 \lesssim z \lesssim 30$, where the IGM is expected to be colder then the CMB. The 21-cm signal can be modelled as a Gaussian absorption profile
\citep[]{bernardi15,presley15}:
\begin{eqnarray}
T_{\rm HI}(\nu_j) = A_{\rm HI} \, \mathrm{e}^{-\frac{(\nu_j - \nu_{\rm HI})^2}{2 \sigma^2_{\rm HI}}},
\label{eq:hi_profile}
\end{eqnarray}
\noindent
where $A_{\rm HI}$, $\nu_{\rm HI}$ and $\sigma_{\rm HI}$ are the
amplitude, peak position and standard deviation of the 21-cm spectrum.
We investigated how well this empirical model reproduces a physical 21-cm spectrum by using the publicly-available code ARES\footnote{https://bitbucket.org/mirochaj/ares} \citep[][]{mirocha12,mirocha14,mirocha15}. Figure~\ref{fig:model_comparison} shows that a Gaussian profile closely resembles the physical reference model defined in \cite{mirocha15} across most of the considered observing band. Deviations between the two models start to become noticeable at high redshift when collisional coupling drives the 21-cm signal negative with respect to the Gaussian model. Although we plan to incorporate physical models in future analyses, the Gaussian profile is sufficiently accurate for the purpose of testing our signal-extraction method and applying it to establish first-order upper limits on the 21-cm signal (Section~\ref{sec:observations}). 

In order to efficiently explore the posterior probability distribution, we use the
sampler \textsc{MultiNEST} \citep[]{feroz08,feroz09}; crucially, it is
an efficient calculator of the Bayesian evidence (with the posterior
samples coming as a by-product) in relatively-low-dimensionality
parameter spaces such as ours, and it robustly uncovers any
degeneracies, skirts, wings or multimodalities in the posterior. We
use an MPI-enabled python wrapper for \textsc{MultiNEST}
\citep{buchner14} that allows a full model fit to be evaluated in just
a few minutes on a typical desktop machine. We have released a python
implementation of our software,
\textsc{hibayes}\footnote{\url{http://github.com/ska-sa/hibayes}
  \citep{ascl_hibayes}.}, that incorporates the models described here,
although the inclusion of different models is straightforward and will
be the goal of future work.

We tested the signal extraction on a simulated case where we considered
a $N = 7$ polynomial foreground model, representing the level of
corruption of the intrinsic sky spectrum due to the primary beam for a
simulated LEDA case \citep{bernardi15}. Such an assumption may be representative of other experiments --- or considered a somewhat pessimistic case.
We adopted the 21-cm model
labelled as `A' in \cite{bernardi15}, which has an amplitude
$A_{\rm HI} = -100$\,mK, a peak frequency $\nu_{\rm HI} = 67$\,MHz, a
width $\sigma_{\rm HI} = 5$\,MHz and similar to the fiducial
model of \cite{pritchard10} and \cite{mirocha15} plotted in Figure~\ref{fig:model_comparison}. We considered a 400-hour integration time
with a 1-MHz channel width and a dual-polarization dipole. We also
assumed the total bandwidth to span the 40--89\,MHz range. These
assumptions, although tuned to the LEDA case, can generally represent
the observing specifications of any ground-based 21-cm global
experiment that targets the pre-reionization era, with the 89-MHz
cutoff being due to the radio frequency interference (RFI) caused by
the radio FM band.

We assumed uniform priors on all the parameters and, in order (solely) to reduce the computing load, we set conservative priors on the 21-cm signal to be $-400 < A_{\rm HI} < 0$~mK, $40 < \nu_{\rm HI} < 89$~MHz and $0 < \sigma_{\rm HI} < 35$~MHz. Whereas the priors on the peak position and width are essentially due to the observational constraints, the amplitude prior can be theoretically motivated by assuming an extreme (and somewhat unlikely) model with no gas heating occurring in the redshift range of interest.

The peak amplitude of the 21-cm signal may be estimated analytically from the
expression for the 21-cm brightness temperature
\citep[e.g.][]{mesinger15}:
\begin{eqnarray}
A_{\rm HI} & \approx & 27 \, x_{\rm HI} \left(1 - \frac{T_\gamma}{T_s} \right) (1 + \delta) \left( \frac{H}{\mathrm{d}v_r/\mathrm{d}r + H} \right)  \nonumber \\
& & \sqrt{\frac{1 + z}{10} \frac{0.15}{\Omega_{\rm M} h^2} } \left( \frac{\Omega_{\rm b} h^2}{0.023} \right) \left( \frac{1-Y_p} {0.75} \right) \, {\rm mK}.
\label{eq:delta_Tb}
\end{eqnarray}
For the global-signal case, we can ignore density fluctuations
(i.e.~set $\delta = 0$), peculiar velocities
(i.e.~set $\mathrm{d}v_r/\mathrm{d}r = 0$) and safely assume the Helium
fraction $Y_p = 0.25$. Assuming the IGM is fully neutral during this epoch (i.e.~$x_{\rm HI} = 1$), Equation~\ref{eq:delta_Tb} becomes
\begin{eqnarray}
A_{\rm HI} \approx 27 \left(1 - \frac{T_\gamma}{T_s} \right)  \sqrt{\frac{1 + z}{10} \frac{0.15}{\Omega_{\rm M}} } \, \frac{\Omega_{\rm b} h}{0.023} \, {\rm mK},
\label{eq:delta_Tb_1}
\end{eqnarray}
where $T_\gamma$ is the CMB temperature, $T_s$ is the spin
temperature, $\Omega_{\rm M} = 0.315$ is the matter density, $\Omega_{\rm b} = 0.049$ is the baryon density and $h \equiv H/(100\,\mathrm{km\,s^{-1}\,Mpc^{-1}}) = 0.673$ is the normalized Hubble parameter \citep{planck15}.

Assuming no gas heating, the gas temperature $T_K$ can be calculated from thermal decoupling (where $T_\gamma = T_K$) following the $(1+z)^2$ adiabatic cooling. Also assuming that the Ly$\alpha$
emission from the first luminous sources is very effective in completely coupling the spin
temperature $T_s$ to the gas temperature, we can write
\begin{eqnarray}
T_s = T_K = T_{\gamma,0} (1 + z_d) \left[ \frac{1 + z}{1 + z_d} \right]^2,
\end{eqnarray}
with $T_{\gamma,0} = 2.73$\,K the CMB temperature at the present time
and $z_d \approx 200$ is the redshift of the thermal decoupling
between the IGM and the CMB. Substituting everything into
Equation~\ref{eq:delta_Tb_1} we obtain
\begin{eqnarray}
A_{\rm HI} \approx 27 \left(1 - \frac{1 + z_d}{1 + z} \right)  \sqrt{ \frac{(1 + z)}{10} \frac{0.15}{\Omega_{\rm M}}} \, \frac{\Omega_{\rm b} h}{0.023} \, {\rm mK},
\label{eq:delta_Tb_2}
\end{eqnarray}
which gives $A_{\rm HI} \approx -380$~mK at $z = 15.7$, corresponding to the lowest redshift of the considered observing band. 

Results of the \textsc{hibayes} fit to the simulated data are shown in
Figures~\ref{fig:triangle_sim_plot}
and~\ref{fig:triangle_sim_zoomin_plot}. Most of the parameters are
well recovered to within the 68-per-cent contours, although some of
the best-fitting values are marginally offset from their true values.
Correlations between some parameters are apparent, although the
one-dimensional marginalized distributions are fairly smooth for all
the parameters, with no evidence for multimodality. Most of the
foreground parameters are very tightly constrained, as are the 21-cm
peak frequency and width. The 21-cm amplitude shows the largest
relative errors --- at the 12-per-cent level --- and noticeable
anti-correlation with the foreground amplitude and slope. Such
anti-correlation can be explained by the degeneracy between these
parameters at the peak frequency $\nu_{\rm HI}$: if the foreground
amplitude is overestimated (underestimated), the 21-cm amplitude will
be underestimated (overestimated) or the foreground slope will be
steeper (flatter), in order to preserve the same observed spectrum
value. We also note that there are correlations between the
higher-order polynomial coefficients, although they can be
disentangled given the high level of sensitivity simulated here. The
results presented here are in agreement with the Fisher matrix
estimates from \cite{bernardi15} and show that, even in the presence
of spectrally-unsmooth foreground emission requiring high-order
polynomials to be modelled, our method is able to extract the
21-cm signal provided sufficient signal-to-noise ratio. Our results
are also broadly consistent with the MCMC analysis presented by
\cite{harker12} and \cite{harker15}, although they used a fairly distinct
frequency band and 21-cm model to ours.

\section{Analysis of LEDA data}
\label{sec:observations}

We next applied our method to preliminary LEDA data. The LEDA instrument is described in detail in upcoming papers
\citep[][]{schinzel16,price2016}; here, we briefly describe the system along with the
observations and data-reduction approach.

LEDA is a sub-instrument of the Long Wavelength Array at the Owens Valley Radio Observatory \citep[LWA-OVRO,][]{hallinan2016}. LWA-OVRO is primarily an all-sky imaging radio interferometer, designed to operate in the frequency range 10--88~MHz. It consists of a `core' of 251 dual-polarization dipole-type antennas within a 200-m diameter, plus an additional 5~`outrigger' antennas, located a few hundred metres from the core, customized for LEDA. In addition, 32~`expansion' antennas are quasi-randomly distributed up to 1500~m from the core.

Each LEDA outrigger antenna is equipped with a receiver board designed for precision radiometry.
Here we present the total-power data taken from a single outrigger antenna, whereas future LEDA analyses will make use of data from all five outrigger antennas, supported by the analysis of
cross-correlations with the core antennas in order to improve the instrument calibration by measuring the antenna primary beam \citep[see the discussion in][]{bernardi15} and ionospheric distortions.

Observations were made during the nights of 2016 February 11 and 12 over a 2-hour period centred at ${\rm LST} = 10^{\rm h} 30^{\rm m}$ when the Galactic centre, Cassiopeia~A and Cygnus~A were near or below the horizon. The antenna total-power data were digitized at a rate of 196.608\,MHz, giving a 0--98.304\,MHz bandwidth, covered by 4096 channels, each of them 24-kHz wide. Data were integrated over 1~second.

We calibrated spectra using a multi-stage approach, as follows. The first calibration stage was a modified version of the three-state switching calibration technique employed by EDGES \citep{rogers12}. The three-state switching removes the effect of time variations in the system gain $G(\nu, t)$ and receiver temperature $T_{\rm{rx}}(\nu,t)$, and imposes an absolute temperature scale on the data. The LEDA outrigger antennas switch between the sky, and two calibration references --- referred to as `hot' and `cold' --- with different noise-equivalent temperatures $T_{\rm{hot}}(\nu,t)$ and $T_{\rm{cold}}(\nu,t)$ respectively. The power measured in each state is then given by:
\begin{align}
P_{\rm{ant}} (\nu,t)	& =G(\nu, t) \, \Delta\nu \, k_{\rm{B}} (T'_{\rm{ant}}(\nu, t) + T_{\rm{rx}}(\nu, t)) \nonumber \\
P_{\rm{hot}} (\nu,t) 	& =G(\nu, t) \, \Delta\nu \, k_{\rm{B}} (T_{\rm{hot}}(\nu, t) + T_{\rm{rx}}(\nu, t)) \nonumber \\
P_{\rm{cold}} (\nu,t)	& =G(\nu, t) \, \Delta\nu \, k_{\rm{B}} (T_{\rm{cold}}(\nu, t) + T_{\rm{rx}}(\nu, t)),
\label{eq:3ss_EDGES}
\end{align}
where $P_{\rm{ant}}$,  $P_{\rm{hot}}$ and $P_{\rm{cold}}$ are the powers for the antenna, hot calibration reference and cold calibration reference states respectively; $T'_{\rm{ant}}$ is the antenna noise-equivalent temperature; $\Delta \nu = 24$~kHz is the channel width and $k_{\rm{B}}$ is the Boltzmann constant. The first-stage calibrated antenna temperature $T'_{\rm{ant}}$ is recovered via
\begin{equation}
T'_{\rm{ant}} (\nu, t) = (T_{\rm{hot}} - T_{\rm{cold}})\frac{P_{\rm{ant}}-P_{\rm{cold}} }{P_{\rm{hot}}-P_{\rm{cold}}}+T_{\rm{cold}}.\label{eq:3ss}
\end{equation}
where $T_{\rm{hot}}$ and $T_{\rm{cold}}$ are measured before the observation and where we drop the explicit dependence on time and frequency on the r.h.s.~of the equation for simplicity.
As the receiver switched between the three states every 5~seconds, with the first second of data in each state blanked, the total on-sky time was eventually 1152~seconds ($\approx 19$~minutes).

The second-stage calibrated antenna temperature $T''_{\rm{ant}}$ is obtained by correcting $T'_{\rm{ant}}$ for the reflection coefficient $\Gamma$:
\begin{equation}
T''_{\rm{ant}} (\nu,t) = T'_{\rm ant} (\nu,t) [1-|\Gamma|^{2} (\nu) ],
\label{eq:vna}
\end{equation}
where $\Gamma (\nu)$ measures the impedance mismatch between the receiver and the antenna and was determined using a vector network analyzer \citep{price2016}.

At this point spectra were flagged for RFI using the \textsc{SumThreshold} algorithm \citep{offringa2010}, then averaged in frequency to achieve a final resolution of 768~kHz. The $40$--$85$~MHz band of interest was subsequently extracted, with a few MHz lost at the upper end of the bandwidth due to filter roll-off.

The final calibration stage was performed using a sky spectrum model $\hat{T}_{\rm{ant}}(\nu,t)$:
\begin{equation}
\hat{T}_{\rm{ant}}(\nu,t)=\frac{\int d\Omega B(\theta,\phi,\nu)T_{\rm{sky}}(\theta,\phi,\nu, t)}{\int d\Omega B(\theta,\phi,\nu)},
\end{equation}
where $B(\theta,\phi,\nu)$ is the antenna beam pattern and $T_{\rm{sky}}(\theta, \phi, \nu, t)$ is the model sky brightness distribution evaluated using the \cite{de-oliveira-costa08} global sky model\footnote{Our python-based implementation that includes observer-centred sky models is available at \url{https://github.com/telegraphic/pygsm} \citep{ascl_pygsm}.}. We used the dipole beam model from \cite{dowell2012}. \cite{dowell16} have recently completed a sky survey covering the frequency range $35-80$~MHz using the LWA and these data may be used to improve the calibration in the future.

The calibrated antenna temperature $T_{\rm ant} (\nu,t)$ measured at time $t$ is then obtained as:
\begin{align}
T_{\rm ant} (\nu,t) & = \frac{\int_T \hat{T}_{\rm ant}(\nu,t') dt'} {\int_T T''_{\rm ant}(\nu,t') dt'} T''_{\rm ant}(\nu,t)\nonumber \\
 & =  \alpha(\nu) T''_{\rm ant}(\nu,t) \label{eqn:calvec},
\end{align}
where the average occurs over the full $T = 2$~hours of observation. The calibration $\alpha(\nu)$ is calculated from observations on 2016 February 11, and then applied to the observations on February 12.

\begin{figure}
\centering
\includegraphics[width=1.0\columnwidth]{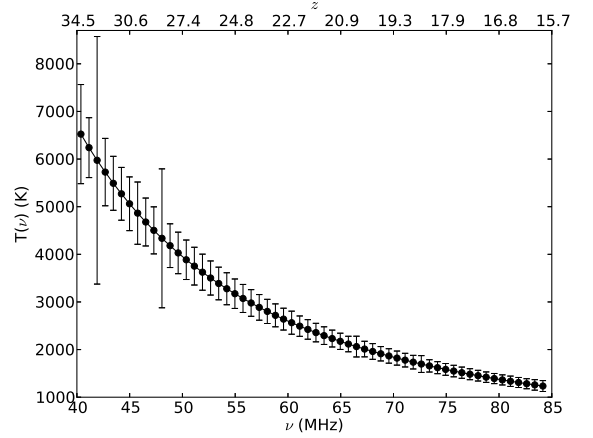}
\caption{Measured sky spectrum for a 2-hour observation ($\approx 19$~minutes effective integration time, 2016 February 12 at $9.5^{\rm h} < {\rm LST} < 11.5^{\rm h}$). Note that the error bars have been inflated by a factor of 1000 in order to make them visible.}\label{fig:observed_spectrum}
\end{figure}
\begin{figure}
\centering
\includegraphics[width=0.9\columnwidth]{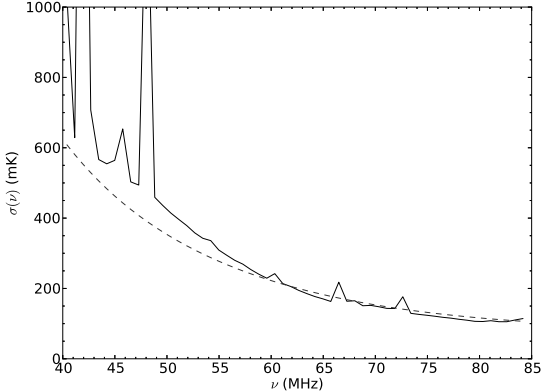}
\caption{Comparison of the estimated thermal noise (dashed line) and the noise measured as the standard deviation of the observed data (solid line).}
\label{fig:rms-comparison}
\end{figure}
The final calibrated sky spectrum $T_{\rm ant} (\nu)$ after averaging
in time is shown in Figure~\ref{fig:observed_spectrum} and becomes the input for \textsc{hibayes}. The thermal noise can be estimated by propagating the uncertainties of Equation~\ref{eq:3ss_EDGES}:
\begin{eqnarray}
\sigma^{2} (\nu) 	& = & \left(\frac{\partial T'_{\rm ant}}{\partial P_{\rm ant}}\right)^{2} (\Delta P_{\rm ant})^2 + \left(\frac{\partial T'_{\rm ant}}{\partial P_{\rm cold}}\right)^{2} (\Delta P_{\rm cold})^2  \nonumber \\
				& + & \left(\frac{\partial T'_{\rm ant}}{\partial P_{\rm hot}}\right)^{2} (\Delta P_{\rm hot})^2.
\label{eq:rms-unc}
\end{eqnarray}
Figure~\ref{fig:rms-comparison} compares the estimated thermal noise with that derived as the standard deviation of the calibrated antenna temperature as a function of time for each frequency channel. The measured and the expected thermal noise levels are consistent above 55\,MHz, whereas the measured noise is higher than expected at lower frequencies. The large spikes below 50\,MHz correlate with known RFI sources, where a larger fraction of data are flagged, causing an effective decrease in integration time.

\begin{figure}
\centering
\includegraphics[width=1.\columnwidth]{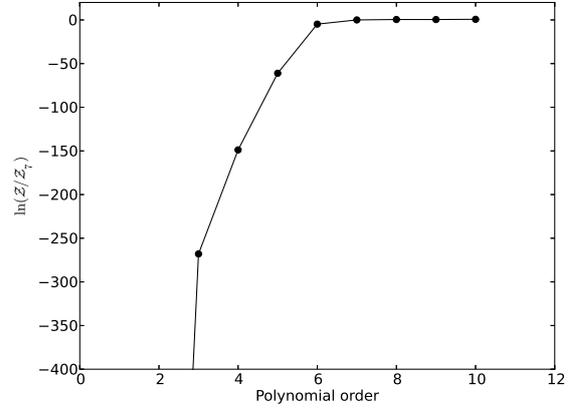}
\caption{Bayesian evidence for foreground models fitted to the LEDA data, relative to the $N=7^{\rm th}$ order polynomial foreground model, as a function of polynomial order $N$. The uncertainties on the evidence are of the same magnitude as the filled circle size.}\label{fig:evidence_plot}
\end{figure}
\begin{figure*}
\centering
\includegraphics[width=1.0\textwidth]{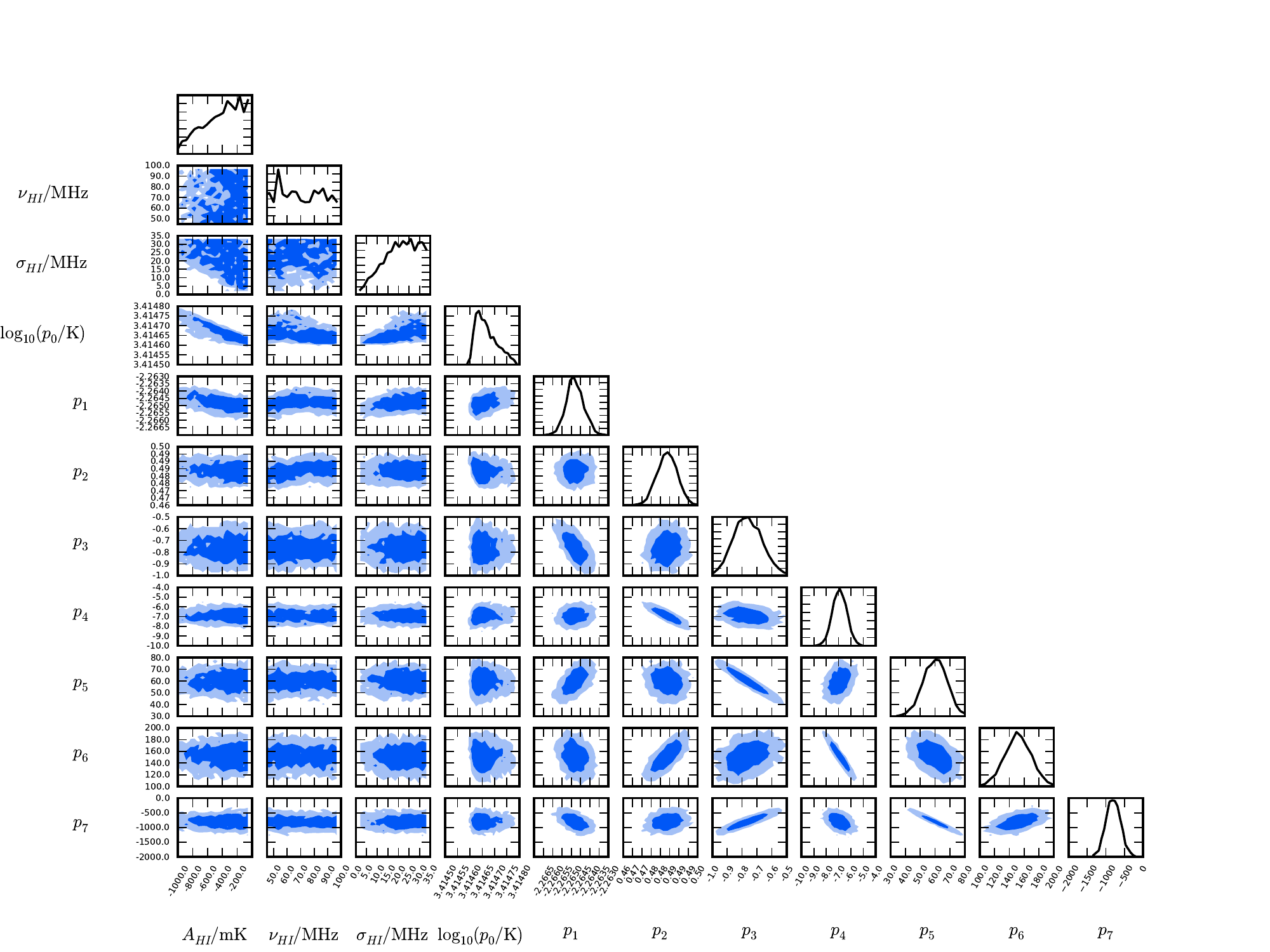}
\caption{Posterior probability distribution, marginalized into one and
  two dimensions, for the $N=7^{\rm th}$ order polynomial foreground
  and 21-cm models, fitted to the LEDA data. The dark and light shaded
  regions indicate the 68- and 95-per-cent confidence
  regions. The marginalized probability distributions are plotted in the $[0,1]$ range.}\label{fig:triangle_plot}
\end{figure*}
In the \textsc{hibayes} analysis, we first looked to confirm the foreground parametrization used in our simulations (section~\ref{sec:bayes}). Following \cite{harker15}, we sought to establish the foreground model by fitting the data with increasing polynomial order assuming that the 21-cm signal is fairly described by Equation~\ref{eq:hi_profile}. We found that the evidence (Figure~\ref{fig:evidence_plot}) increases sharply as a function of polynomial order until $N=6$, after which it starts to flatten.
According to the scale of \cite{jeffreys39}, the $N=7$ model is still decisively (odds $> 100:1$) preferred over the $N=6$ model, whereas the $N=8$ model is disfavoured (negative odds) over the $N=7$ model. In practice the evidence remains essentially flat as the polynomial increases beyond the $N=7^{\rm th}$ order and small (positive or negative) variations are likely due to sampling accuracy. We therefore fitted the data using an evidence-motivated model that includes the 21-cm signal and a $N=7^{\rm th}$ order polynomial foreground model and we used the measured noise as a function of frequency. We emphasize here that the chosen $N=7^{\rm th}$ order polynomial is not intended to represent the spectral structure of the intrinsic sky emission but rather the `observed foregrounds', i.e.~the convolution of the intrinsic sky emission with the instrumental response. It is also worth noticing that the evidence here favours a model that is in fair agreement with earlier LEDA simulations presented in \cite{bernardi15}. In a future work we will investigate the possibility of parameterizing the instrument and the sky emission
separately and of using the evidence to indicate the best choice of sub-models, rather than assuming a combined parameterization as here.\\

After having established the foreground model, we set the priors on the width of the 21-cm signal to be the same used for simulations as they encompass the full breadth of theoretical predictions. We set uniform priors on the 21-cm peak position to be $50 < \nu_{\rm HI} < 100$~MHz as such a range brackets both models with the most extreme star-formation efficiency --- which would shift the peak at low frequencies --- and with the most extreme X-ray efficiency --- which would shift the peak at high frequencies \citep[][]{pritchard10,mirocha15}. We relaxed the constraints on the depth of the 21-cm peak amplitude that we used for the simulated case because we seek to derive data-driven upper limits even in the case of models that are disfavoured by theory. We still assumed that the 21-cm signal cannot be positive, i.e.~that $T_s = T_k < T_\gamma$, which is accepted in any model for the redshift range considered here.

We ran different chains with decreasing lower bounds of the $A_{\rm HI}$ prior. In the initial case we set $-380 < A_{\rm HI} < 0$~mK and found that the whole prior range is within the 95-per-cent confidence region. We found that the two-dimensional posterior distributions for both the 21-cm amplitude and the width showed monotonically-decreasing profiles with increasing prior range until an area of the prior range is clearly excluded at a confidence level greater than 95-per-cent for $-1000 < A_{\rm HI} < 0$~mK.
The posterior probability distribution for this final run is displayed in Figure~\ref{fig:triangle_plot}.

The foreground parameters are very well constrained and all their marginalized, one-dimensional distributions are Gaussian-like, similar to the simulated case, with the exception of the foreground amplitude $p_0$, whose marginalized posterior is slightly asymmetric.
The best-fit foreground parameters are fairly different from the simulated case apart from the first two coefficients that indicate the foreground amplitude at 60\,MHz and its power-law slope. This is not unexpected as higher-order polynomials are most likely compensating for limitations in the instrumental calibration that were not included in the simulations (e.g.~errors on the reflection coefficient or calibration load). It is interesting, however, that the best-fit foreground coefficients show correlations between foreground parameters similar, albeit at a qualitative level, to the simulated case, for example between the $p_3$ and $p_5$, and $p_5$ and $p_7$, coefficients.

The 68- and 95-per-cent confidence contours of the 21-cm parameters are fairly different and, essentially, identify upper limit regions, as expected given the noise levels. Whereas no constraints can be placed on the peak position $\nu_{\rm HI}$ within the prior range, bright, narrow 21-cm Gaussian profiles are disfavoured by the data. Quantitatively, we constrain $A_{\rm HI} > -890$~mK and $\sigma_{\rm HI} > 6.5$~MHz at the 95-per-cent level in the $13.2 < z < 27.4$ range: this amplitude limit is only a factor of $\approx 2.5$ away from constraining the extreme model with no heating described in Section~\ref{sec:bayes}.

\begin{figure}
\includegraphics[width=1.\columnwidth]{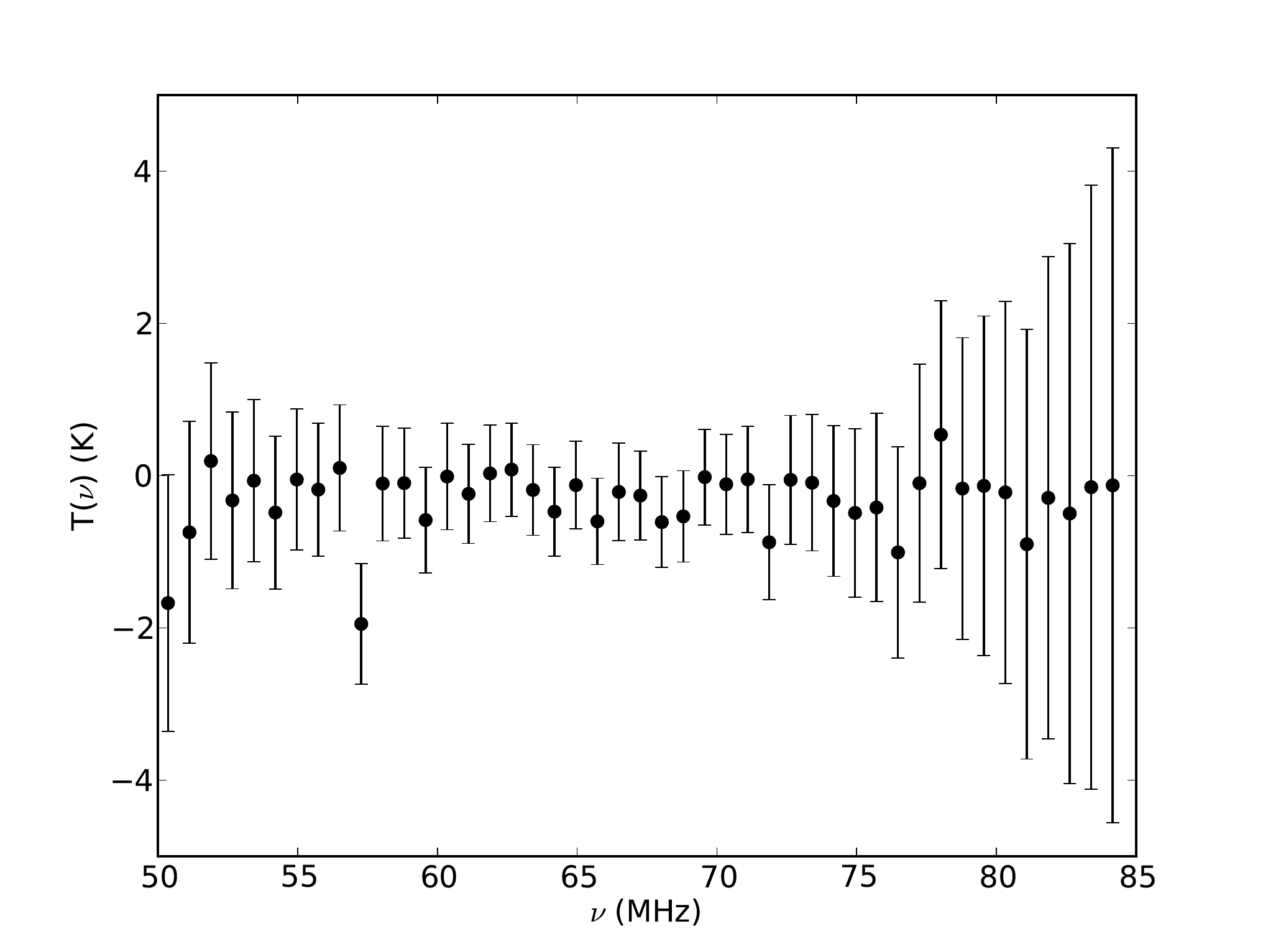}
\caption{Residual spectrum after subtraction of the best-fit, maximum \textit{a posteriori} foreground model. Error bars are plotted at the $2$-$\sigma$ confidence level and include both the measured and the best-fit parameter uncertainties. All the data points but one at 57.7~MHz are compatible with zero.}
\label{fig:residuals}
\end{figure}

In the absence of a detection, the root-mean-square of the residuals is a metric often used in observations --- although less statistically rigorous than limits derived directly from the posterior probability distribution. Figure~\ref{fig:residuals} shows the residual spectrum after subtraction of the best-fit, maximum \textit{a posteriori} foreground model. We find a 470~mK rms residual over the whole redshift range, that, if considered a proxy for the 68-per-cent confidence level, is approximately consistent with the \textsc{hibayes} constraints on the 21-cm peak amplitude.

\section{Summary and Conclusions}
\label{sec:conclusions}
We have presented a fully-Bayesian algorithm for simultaneously fitting the global 21-cm signal in the presence of sky foregrounds. Our algorithm capitalizes on the Bayesian evidence's Occam's razor effect for model selection, with posterior probability distributions coming as a by-product.

We tested the method on simulated data and showed that, assuming a 7$^{\rm th}$-order polynomial foreground spectrum, the 21-cm global signal --- parameterized as a Gaussian absorption profile --- can be strongly constrained with a 400-hour integration time for a LEDA-like observing setup. This result more quantitatively confirms the Fisher matrix analysis previously carried out by \cite{bernardi15}. Although here we presented a specific application for the 21-cm signal from the Cosmic Dawn, the code can easily be extended to the full redshift range of interest for global-signal measurements. The code \citep{ascl_hibayes} is publicly available at \url{http://github.com/ska-sa/hibayes}.

We applied the method to observations in order to derive upper limits on the 21-cm signal from the Cosmic Dawn. We showed that the Bayesian evidence can guide the choice of foreground model \citep{harker15}, with a maximum for a $7^{\rm th}$-order polynomial in the present case. We emphasize that such a model reflects the combination of the intrinsic foregrounds and the spectral structure introduced by the instrument. In this respect, the evidence does not yet constrain the intrinsic foreground spectrum as suggested by \cite{harker15} and future work will be dedicated to incorporating both the intrinsic sky and instrument models in the analysis and placing constraints on the intrinsic foreground spectrum.

The best-fit foreground parameters are very well constrained; in particular we derive a spectral index for the diffuse Galactic emission $\beta (\equiv p_1) = 2.27 \pm 0.04$. This value is consistent with early measurements of the Galactic radio background at 81.5\,MHz by \cite{bridle67}, but is noticeably flatter than what was measured at 150~MHz by \cite{rogers08}. This possible flattening of the spectral index may be good news for foreground subtraction for future 21-cm (global and interferometric) observations targeting the pre-reionization epoch.

\cite{voytek14} report the only other broadband measurements at these frequencies. A direct comparison with their results is not straightforward as they do not report either r.m.s.~residuals or direct upper limits on 21-cm parameters. We note, however, that our best-fit spectral index is consistent with theirs, although their spectrum normalization is about 40-per-cent greater than what we report here. Our measurements, however, are within 10-per-cent of the carefully absolutely-calibrated Galactic spectrum measured by EDGES at 150~MHz \citep{rogers08} once it is scaled down to 60~MHz using the EDGES spectral index. We therefore believe our absolute flux density scale to be appropriate and its uncertainty to be negligible at the present level of sensitivity.

Our analysis constrains the 21-cm signal amplitude and width to be $-890 < A_{\rm HI} < 0$~mK and $\sigma_{\rm HI} > 6.5$~MHz respectively at the 95-per-cent confidence level in the $13.2 < z < 27.4$ ($100 > \nu > 50$~MHz) range. Note that the constraint on $\sigma_{\rm HI}$ corresponds to a redshift width $\Delta z\approx 1.9$ at redshift $z\simeq 20$.
Our results are the tightest upper limits on the 21-cm signal from the Cosmic Dawn to date and are encouraging in terms of achieving a factor of a few improvement in the sensitivity necessary to start placing significant constraints on structure prior reionization and on the thermal history of the IGM and the related sources of heating.

\section*{Acknowledgments}
\label{sec:acknowledgments}

We thank an anonymous referee for helpful comments that considerably improved the manuscript. GB thanks Judd Bowman, Andrea Ferrara, Adrian Liu and Aaron Ewall-Wice for useful inputs and comments on this work and Jordan Mirocha for help with ARES. The LEDA experiment is supported by NSF grants AST/1106059 and PHY/0835713. JZ gratefully acknowledges a South Africa National Research Foundation Square Kilometre Array Research Fellowship. This research was supported by the Munich Institute for Astro- and Particle Physics (MIAPP) of the DFG cluster of excellence `Origin and Structure of the Universe'. With the support of the Ministry of Foreign Affairs and International Cooperation, Directorate General for the Country Promotion (Bilateral Grant Agreement ZA14GR02 - Mapping the Universe on the Pathway to SKA).

\bibliography{ledabayes}\label{lastpage}
\bibliographystyle{mnras}
\bsp

\onecolumn

\end{document}